\documentclass{article}

\usepackage{arxiv}
\usepackage{graphicx}
\usepackage{caption}
\usepackage{subcaption}
\usepackage[utf8]{inputenc} 
\usepackage[T1]{fontenc}    
\usepackage{hyperref}       
\usepackage{url}            
\usepackage{booktabs}       
\usepackage{amsfonts}       
\usepackage{nicefrac}       
\usepackage{microtype}      
\usepackage{lipsum}

\title{Prognostic power of texture based morphological operations in a radiomics study for lung cancer}

\author{
  Paul Desbordes \\
  ICTEAM\\
  Université catholique de Louvain\\
  Louvain-la-Neuve, Belgium\\
  \texttt{paul.desbordes@uclouvain.be} \\
  \And
  Diksha \\
  Indian Institute of Technology Ropar\\
  Punjab, India\\
  \texttt{2016eeb1074@iitrpr.ac.in} \\
  \And
  Benoit Macq \\
  ICTEAM\\
  Université catholique de Louvain\\
  Louvain-la-Neuve, Belgium\\
  \texttt{benoit.macq@uclouvain.be} \\
}

\begin{document}
\maketitle

\begin{abstract}
The importance of radiomics features for predicting patient outcome is now well-established. Early study of prognostic features can lead to a more efficient treatment personalisation. For this reason new radiomics features obtained through mathematical morphology-based operations are proposed. Their study is conducted on an open database of patients suffering from Nonsmall Cells Lung Carcinoma (NSCLC). The tumor features  are extracted from the CT images and analyzed via PCA and a Kaplan-Meier survival analysis in order to select the most relevant ones. Among the 1,589 studied features, 32 are found relevant to predict patient survival: 27 classical radiomics features and five MM features (including both granularity and morphological covariance features). These features will contribute towards the prognostic models, and eventually to clinical decision making and the course of treatment for patients.
\end{abstract}


\section{Introduction}
Radiomics is a fast-growing concept that aims for high-throughput extraction and analysis of large amounts of quantitative features from clinical images~\cite{lambin2012}. Advanced analysis can reveal the prognostic and the predictive power of features which helps avoid invasive exams while improving the treatment decisions and the likely course of the disease for the patient. Several articles  can already be found in the literature . In this article, we focus on lung cancer examined by CT images which are particularly interesting to reveal the high-density contrast between tumor and lungs, as well as the intra-tumor textures and the tumor shape.

In~\cite{aerts2014}, Aerts et al. studied the relevance of 440 radiomics features extracted from CT of 1,019 patients with head and neck cancer (H\&N) and NonSmall Cell Lung Carcinoma (NSCLC). It revealed that a large number of them have a prognostic power ($p>5\%$). By combining four of them, the prognostic performance was slightly higher than TNM stage or tumor volume that are the references in this kind of study. This signature, capturing intra-tumor heterogeneity, was validated latter by Leijenaar et al. ~\cite{leijenaar2015} on a cohort of patients suffering from oro-pharyngeal squamous cell carcinoma. In~\cite{aerts2016}, Aerts et al. investigated the relevance of CT radiomics features to predict mutations status in NSCLC. From their study, one radiomics feature (laws-energy) appeared to be significantly predictive of this mutation (AUC$=0.67$, $p=0.03$) whereas the tumor volume is not ($p>5\%$). Other studies have shown the importance of radiomics features in predicting pathological response prior to surgery~\cite{coroller2017}, for response assessment following ablative radiotherapy for early stage NSCLC~\cite{mattonen2016} and for estimating disease-free survival in patients with early-stage NSCLC~\cite{huang2016}.

The Image Biomarker Standardisation Initiative (IBSI) proposed recommendations to standardise the radiomics process ~\cite{zwanenburg2016}, especially promoting features with best reproducibility. Despite a controlled protocol, conclusions about radiomics features varied from a study to another. Ger et al.~\cite{ger2019} retrospectively studied the prognostic power PET/CT features from a large cohort of H\&N cancer patients. They found a higher Area Under the Curve (AUC) when studying tumor volume alone rather than its combination radiomics features meaning they are not consistently associated with survival in CT or PET images of H\&N patients, even with the same protocol.

With the aim of finding new relevant features quantifying medical images in a reproducibility way, we propose to study Mathematical Morphology (MM) based features. MM provides a very rich variety of texture descriptors that benefit from its shape-based nature and its capacity to exploit the spatial relationships among pixels~\cite{aptoula2011}. High classification performances have been observed in various fields of image processing such as automatic recognition of cancerous tissues based on cells analysis using microscopic images~\cite{thiran1996}, image enhancement~\cite{bai2015} and extraction of numerical features from solar images~\cite{stenning2013}. Aptoula et al.~\cite{aptoula2011} suggested that these interesting results are obtained thanks to the capacity of morphological series to capture higher order properties of spatial random processes without being limited by the theoretical bounds imposed on alternative approaches. Despite these interesting results, MM features have never been studied, to our knowledge, in a radiomics context.

In this paper we propose to study the prognostic power of features extracted with MM tools. For that, a radiomics study is performed using an open access medical images dataset composed of patients suffering from lung cancer~\cite{aerts2015}. A Kaplan-Meier survival analysis is performed to study the features relevance to predict Overall Survival (OS). Furthermore, MM based features are compared with classical radiomics (CR) features (first-order, shape-based and textures) to check if these new features bring new information.

\section{Materials and methods}
\subsection{Patient population}
The dataset retrospectively studied contains a total of 422 patients suffering from NSCLC ~\cite{aerts2015}. The inclusion criteria for our study are the availability of a chest CT exam performed before the treatment, a manual segmentation of the tumor performed by an expert and the integrity of the clinical data. As a result, the dataset shrank from 422 to 297 patients. Several clinical and demographic features of these patients are summarized in Table~\ref{table:featData}.

\begin{table}[h]
    \caption{List of clinical and demographic features of the 297 NSCLC patients included in our study.}
    \centering
    \begin{tabular}{ll}
    \toprule
    Features & Number of patients\\ \hline
    
    \textit{Demographic} & \\
    \hspace{1mm} Patient’s age (years) \\
    \hspace{2mm} Median (range) & 68.3 (42.5 - 91.7) \\
    \hspace{2mm} Mean (standard deviation) & 68.0 (10.0) \\
    \hspace{1mm} Patient’s gender & \\
    \hspace{2mm} Male & 211 (71\%) \\
    \hspace{2mm} Female & 86 (29\%) \\
    \hline
    
    \textit{Clinical} & \\
    \hspace{1mm} Histology & \\
    \hspace{2mm} Adenocarcinoma & 30 (10.1\%) \\
    \hspace{2mm} Not otherwise specified & 50 (16.8\%) \\
    \hspace{2mm} Large cells & 96 (32.3\%) \\
    \hspace{2mm} Squamous cell Carcinoma & 85 (28.6\%) \\
    \hspace{2mm} Unknown & 36 (12.1\%) \\
    \hspace{1mm} TNM clinical stage & \\
    \hspace{2mm} I & 76 (25.6\%) \\
    \hspace{2mm} II & 25 (8.4\%) \\
    \hspace{2mm} III (a \& b) & 195 (65.7\%) \\
    \hspace{2mm} Unknown & 1 (0.3\%) \\
    \hline
    
    \textit{Outcomes} & \\
    \hspace{1mm} Follow-up (months) & \\ 
    \hspace{2mm} Median (range) & 16.1 (0.8 - 71) \\
    \hspace{2mm} Mean (standard deviation) & 19.5 (14.9) \\
    \hspace{1mm} Last news survival & \\
    \hspace{2mm} Alive & 99 (33.3\%) \\
    \hspace{2mm} Dead & 198 (66.7\%) \\
    \hspace{1mm} Two-years survival & \\
    \hspace{2mm} Alive & 128 (43.1\%) \\
    \hspace{2mm} Dead & 169 (56.9\%) \\
    \bottomrule
    \end{tabular}
    \label{table:featData}
\end{table}

CT images were acquired using different devices from Siemens Healthcare (Biograph 40, Sensation 10, Sensation 16 and Sensation Open) and using different parameters. Because of the variability of voxels resolution and according to the ISBI recommendations~\cite{zwanenburg2016}, a resampling step is performed to reach a spacing of 1$\times$1$\times$1~mm\textsuperscript{3}, which improves data homogeneity.

\subsection{Feature extraction}
The following workflow is performed to extract features from the CTs. The tumor volume is firstly manually segmented by experts leading to a mean tumor volume of $26.6 \pm 35.2$~cm\textsuperscript{3} (range: $0.2-263.7$~cm\textsuperscript{3}). From this Volume-Of-Interest (VOI), two categories of features are extracted : CR and MM based features. 
The CR features are extracted following the ISBI protocol leading to ten first-order features, one shape feature and 37 texture features coming from four different texture matrices (Gray-Level Cooccurrence Matrix (GLCM), Gray-Level Run-Length Matrix (GLRLM), Gray-Level Size-Zone Matrix (GLSZM) and Gray-Level Difference Matrix (GLDM)). GLCM and GLRLM are computed in all the thirteen 3D directions (separated by 45\textdegree) within the neighborhood of Chebyshev with a distance of one. Features are then extracted from the resulting mean matrix. 

For the MM based features, the methodology proposed by~\cite{aptoula2011} is used. Two kinds of MM features are extracted: granularity and morphological covariance (including granularity and covariance moments). Ten successive iterations for each MM granularity feature and 130 for each MM covariance feature are performed, giving us a total of 1,550 MM based features. Here again the Chebyshev thirteen unique directions are used with a distance of one. In Table~\ref{table:feats} are listed all the image features used in this study (49 CR features and 1550 MM features).

\begin{table}[!ht]
    \centering
    \caption{List of the studied image features.}
    \begin{tabular}{ll}
    \toprule
    Type of features & Features\\ 
    \hline
    First-order & Volume, Sum of intensities (I\textsubscript{sum}), \\
    & I\textsubscript{max}, I\textsubscript{min}, I\textsubscript{mean}, I\textsubscript{std}, I\textsubscript{cov}, \\
    & Skewness, Kurtosis, Energy, Entropy \\

\hline
Shape based & Sphericity \\

\hline
Radiomics textures & GLCM: Variance, Energy, Entropy, Correlation, \\
& \hspace{1mm}  Dissimilarity, Contrast, Homogeneity, Inverse Differential \\
& \hspace{1mm} Moment (IDM), Cluster shade, Cluster tendency \\
& GLRLM: Short Run Emphasis (SRE), Long Run \\
& \hspace{1mm} Emphasis (LRE), Low Gray-level Run Emphasis (LGRE), \\
& \hspace{1mm} High Gray-level Run Emphasis (HGRE), Short Run Low \\
& \hspace{1mm} Gray-level Emphasis (SRLGE), Long Run Low \\
& \hspace{1mm} Gray-level Emphasis (LRLGE), Short Run High \\
& \hspace{1mm} Gray-level Emphasis (SRHGE), Long Run High \\
& \hspace{1mm} Gray-level Emphasis (LRHGE), Run Percentage (RP), \\
& \hspace{1mm} Gray-level Non-Uniformity (GLNUr), Run Length Non \\
& \hspace{1mm} Uniformity (RLNU) \\
& GLSZM: Short Zone Emphasis (SZE), Long Zone Emphasis\\
& \hspace{1mm} (LZE), Low Gray-level Zone Emphasis (LGZE), \\
& \hspace{1mm} High Gray-level Zone Emphasis (HGZE), Short \\
& \hspace{1mm} Zone Low Gray-level Emphasis (SZLGE), Long Zone \\
& \hspace{1mm} Low Gray-level Emphasis (LZLGE), Short Zone High \\
& \hspace{1mm} Gray-level Emphasis (SZHGE), Long Zone High \\
& \hspace{1mm} Gray-level Emphasis (LZHGE), Zone Percentage (ZP), \\
& \hspace{1mm} Gray-level Non-Uniformity (GLNUz), Zone Length Non \\
& \hspace{1mm} Uniformity (ZLNU) \\
& GLDM: Coarseness, Contrast, Busyness, Complexity, \\
& \hspace{1mm} Strength \\

\hline
MM Granularity & Volume (Sum of intensities) \\
 & $\eta\textsubscript{300}, \eta\textsubscript{030}, \eta\textsubscript{003}$ \\
\hline
MM Granularity Moments & Maximum, Standard deviation, Covariance, \\
& Skewness, Kurtosis, Energy, Entropy
\\
\hline
MM Covariance & Volume (Sum of intensities) \\
 & $\eta\textsubscript{300}, \eta\textsubscript{030}, \eta\textsubscript{003}$ \\
\hline
MM Covariance Moments & Maximum, Standard deviation, Covariance, \\
 & Skewness, Kurtosis, Energy, Entropy \\
\bottomrule
\end{tabular}
\label{table:feats}
\end{table}

\subsection{Statistical analysis}
To keep uncorrelated features and eliminate redundant ones, a Spearman's rank correlation analysis is performed as a first step. The 1,589 features are compared one by one to bring out non-linear relationships. They are considered as significant if the absolute value of the Spearman’s correlation coefficient ($|\rho|$) is higher or equal to 0.8 with a $p$-value smaller than 5\%~\cite{desbordes2017}.

Next, Principal Component Analysis (PCA) is used on the MM features to reduce dimensionality by preserving most of the valuable information~\cite{bharati2001}. PCA allows us to evaluate the intra-class variability along the direction of the largest variance in the feature space. For each of the 22 MM features in Table~\ref{table:feats}, we retain 5 PCA components, since they contain most of the information. This step leads to a selection of 49 CR and 110 MM features, for a total of 159 post-PCA features. 

To assess the prognostic value of these features, a univariate Kaplan-Meier analysis is performed to estimate the survival distribution. The OS is calculated from the date of the initial diagnosis to the date of death or to the end of the follow-up period. Patients who are alive are censored at the time of the last recording. The association between OS and each feature is performed after a dichotomization process. The most discriminating cut-off value allowing the differentiation of the two groups of patients is selected using the Receiver Operating Characteristic (ROC) methodology. ROC curves led to AUC, Sensitivity (Se) and Specificity (Sp). The prognostic value of each feature is assessed using the log-rank test where a $p$-value less than 5\% is considered to be statistically significant. The Kaplan-Meier analysis leads to median survival, percentage of deaths in each group and Hazard Ratio (HR).

To avoid false conclusions, appropriate statistical corrections for the type-I errors are performed according to Chalkidou et al.~\cite{chalkidou2015}. For each $p$-value calculated, a Benjamini-Hochberg correction for multiple hypotheses testing is applied. Furthermore, a correction of the minimal $p$-values obtained from the optimum cut-off approach is performed using the Altman formula~\cite{altman1994}.

\section{Results}
\subsection{Correlation analysis}
\begin{figure}
     \centering
     \begin{subfigure}[b]{.4\textwidth}
         \centering
         \includegraphics[width=7cm]{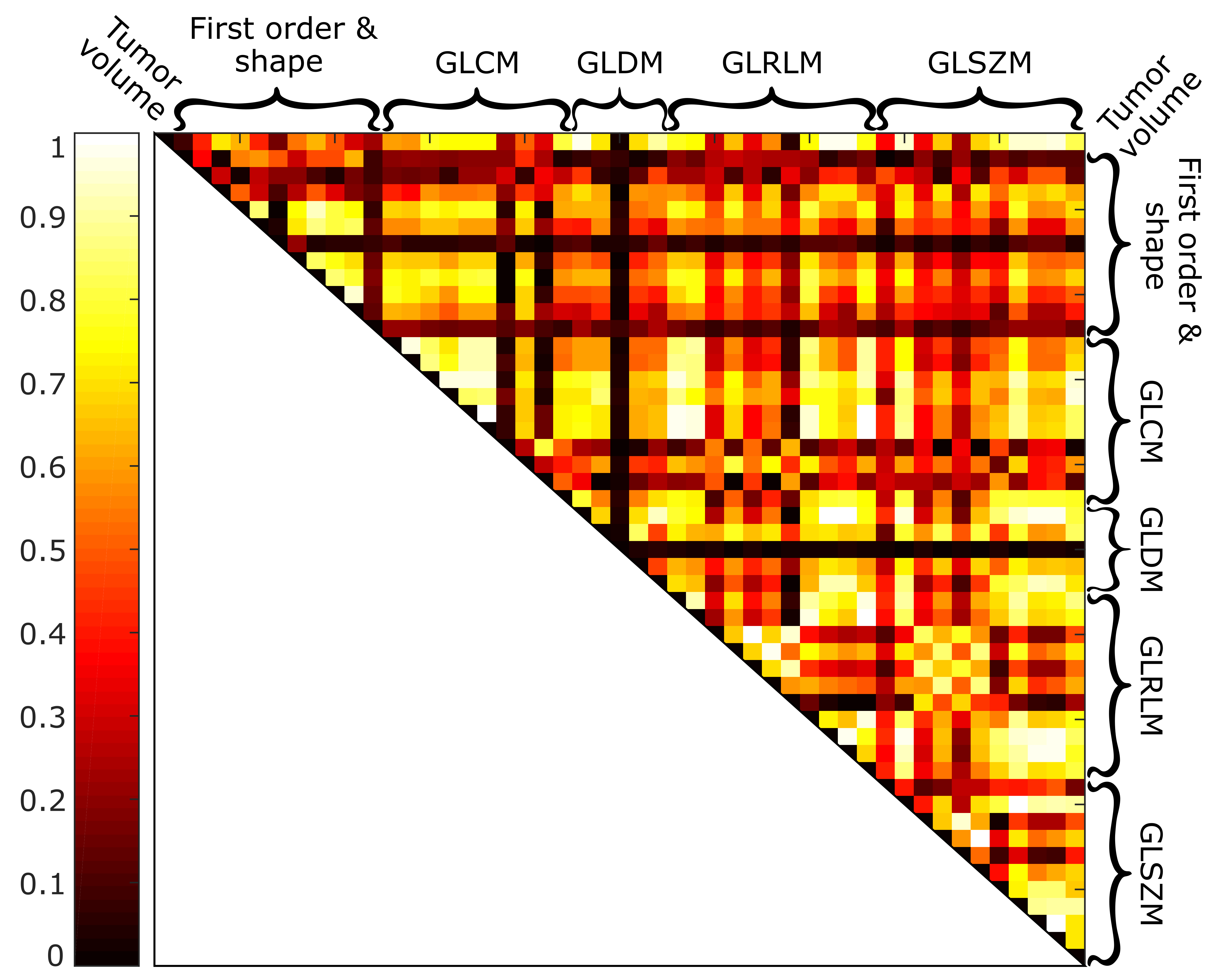}
         \caption{}
         \label{fig:correl_radiom}
     \end{subfigure}
     \hfill
     \begin{subfigure}[b]{.4\textwidth}
         \centering
         \includegraphics[width=7.5cm]{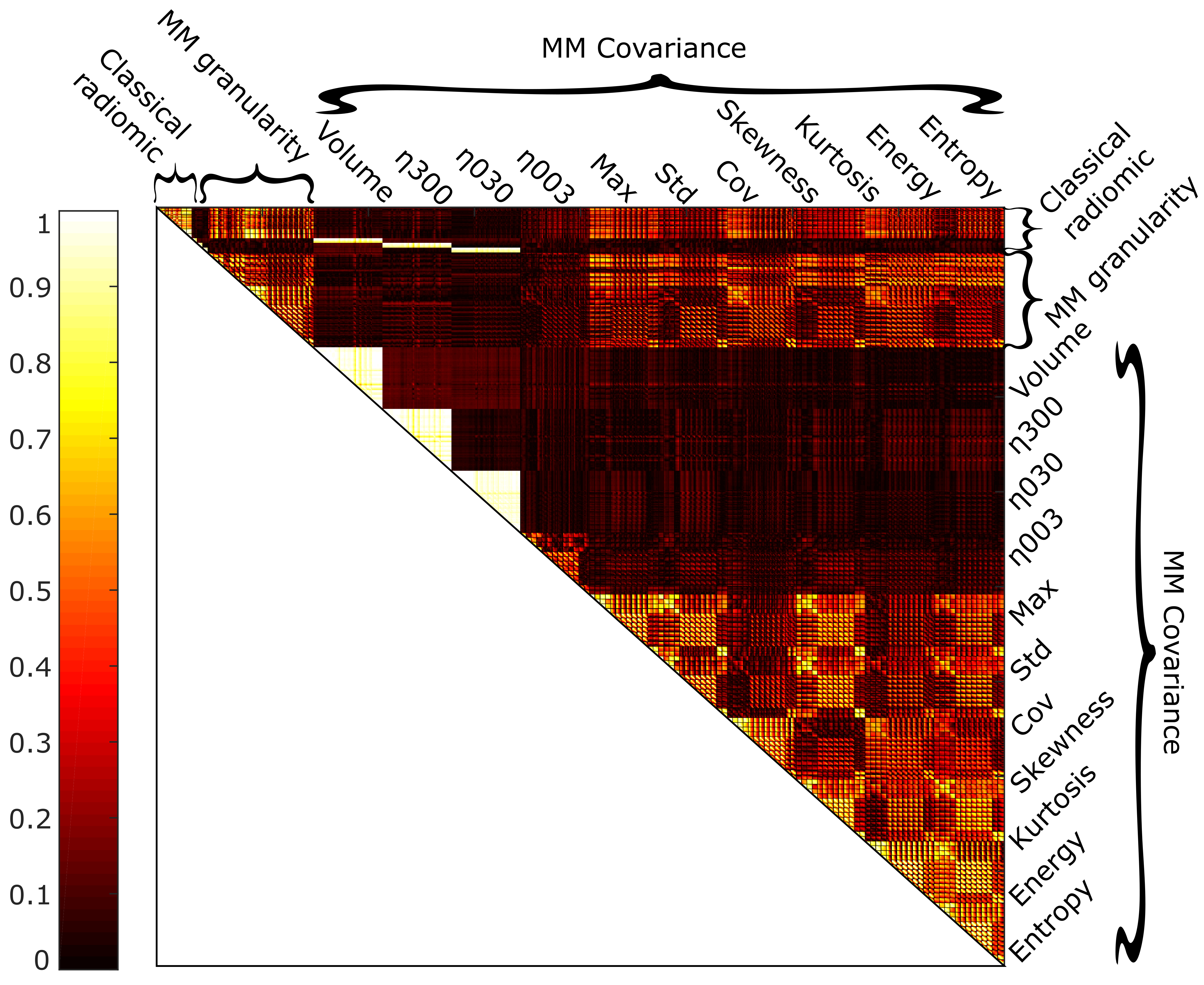}
         \caption{}
         \label{fig:correl_all}
     \end{subfigure}
    \caption{Matrices of Spearman's correlation absolute coefficient (\boldmath$|\rho|$) (a) of the 49 classical radiomics features and (b) of all of the 1,589 studied features}
    \label{fig:correl}
\end{figure}

Results of the correlation study are represented Figure~\ref{fig:correl}. A first focus on the 49 CR features is performed. Because, no correlation pattern appears, we chose to continue to study all of them without any PCA reduction. 

In Figure~\ref{fig:correl_all} are represented the correlations between the 1,589 studied features (CR and MM features). Because there is no correlation pattern between CR, granularity and morphological covariance features we chose to deal with them separately. Among the different iterations of MM based features, significant correlations can be seen per metric. Correlations can also be noticed in some regions within the granularity and morphological covariance features showing that these features are intra-correlated. 

\begin{figure}
\centering
\includegraphics[width=6cm, height=6cm]{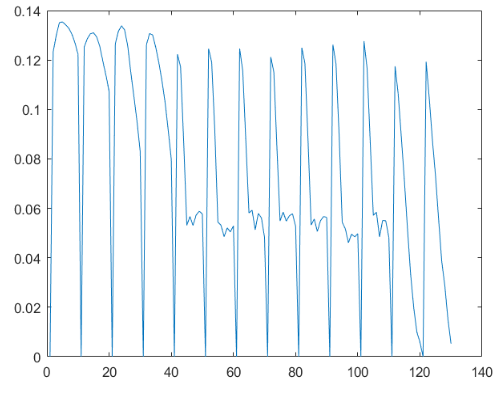}
\caption{PCA coefficient used for the volume-PCA1 from the MM covariance.}
\label{fig:pca}
\end{figure}

Due to this redundancy and non-sparsity of features, PCA is used to get non-correlated features. For both the MM granularity features, the PCA coefficients are similar for all the 10 iterations, showing that there is no preference between all the initial features. However, for the MM covariance features, we observed a cycle of spikes for every 10 iterations for all the 130 iterations. This cycle is illustrated in Figure~\ref{fig:pca} for the case of volume-PCA1.

\subsection{Statistical analysis}
The univariate Kaplan-Meier survival analysis results for relevant features are given Table~\ref{table:proStu}. After our experiments, we found that 32 features among the 159 studied are relevant to predict the survival at two years: 27 CR features and 5 MM features. 

\begin{table}
\centering
\caption{
List of the features with a significant prognostic power ($\alpha < 5\%$) according to the Kaplan-Meier analysis for patients with lung cancer. Sensitivity (Se), Specificity (Sp), Area Under Curves (AUC), Hazard Ratio (HR) and $p$-value for relevant prognostic features. CI: Confidence Interval.}
\begin{tabular}{llccccc}
\toprule
Rank & Feature & Se & Sp & AUC & HR (95\% CI) & $p$ \\
\hline
1 & GLNUz (GLSZM) & 0.65 & 0.61 & 0.618 & 2.00 (1.51-2.66) & 3.48e-5 \\
2 & Coarseness (GLDM) & 0.74 & 0.50 & 0.585 & 2.06 (1.56-2.73) & 3.48e-5 \\
3 & RP (GLRLM) & 0.73 & 0.50 & 0.570 & 2.08 (1.57-2.74) & 3.91e-5 \\
4 & Tumor volume (mL) & 0.67 & 0.54 & 0.565 & 2.02 (1.53-2.67) & 4.24e-5 \\
5 & Complexity (GLDM) & 0.50 & 0.78 & 0.618 & 1.94 (1.43-2.63) & 7.55e-5 \\
6 & GLNUr (GLRLM) & 0.73 & 0.53 & 0.597 & 2.01 (1.52-2.66) & 7.65e-5 \\
7 & Strength (GLDM) & 0.70 & 0.56 & 0.633 & 1.93 (1.46-2.55) & 1.88e-4 \\
8 & LRHGE (GLRLM) & 0.43 & 0.74 & 0.568 & 1.89 (1.39-2.55) & 2.27e-4 \\
9 & LZHGE (GLSZM) & 0.81 & 0.37 & 0.591 & 2.02 (1.52-2.69) & 3.75e-4 \\
10 & ZLNU (GLSZM) & 0.76 & 0.46 & 0.597 & 1.92 (1.45-2.54) & 5.06e-4 \\
11 & Volume-PCA1 (MM Granularity) & 0.63 & 0.63 & 0.631 & 1.84 (1.39-2.44) & 5.43e-4 \\
12 & LZE (GLSZM) & 0.63 & 0.54 & 0.577 & 1.84 (1.39-2.44) & 6.77e-4 \\
13 & SRE (GLRLM) & 0.81 & 0.32 & 0.558 & 1.97 (1.48-2.62) & 9.88e-4 \\
14 & LRE (GLRLM) & 0.30 & 0.84 & 0.544 & 1.85 (1.32-2.58) & 1.02e-3 \\
15 & Kurtosis-PCA3 (MM covariance) & 0.70 & 0.58 & 0.618 & 1.81 (1.37-2.39) & 1.69e-3 \\
16 & Energy-PCA2 (MM covariance) & 0.81 & 0.40 & 0.587 & 1.89 (1.42-2.50) & 1.73e-3 \\
17 & I\textsubscript{mean} & 0.57 & 0.63 & 0.596 & 1.76 (1.33-2.34) & 1.76e-3 \\
18 & HGRE (GLRLM) & 0.70 & 0.49 & 0.593 & 1.79 (1.35-2.36) & 2.13e-3 \\
19 & Entropy (GLCM) & 0.75 & 0.44 & 0.602 & 1.81 (1.36-2.39) & 2.51e-3 \\
20 & Kurtosis & 0.60 & 0.65 & 0.613 & 1.73 (1.31-2.29) & 2.83e-3 \\
21 & SRHGE (GLRLM) & 0.83 & 0.35 & 0.588 & 1.86 (1.40-2.48) & 3.22e-3 \\
22 & RLNUr (GLRLM) & 0.31 & 0.80 & 0.542 & 1.76 (1.26-2.44) & 3.71e-3 \\
23 & Correlation (GLCM) & 0.73 & 0.42 & 0.549 & 1.75 (1.32-2.31) & 5.96e-3 \\
24 & Contrast (GLDM) & 0.64 & 0.57 & 0.601 & 1.69 (1.28-2.24) & 6.10e-3 \\
25 & Dissimilarity (GLCM) & 0.36 & 0.81 & 0.596 & 1.69 (1.23-2.33) & 8.04e-3 \\
26 & IDM (GLCM) & 0.38 & 0.78 & 0.577 & 1.68 (1.23-2.29) & 8.37e-3 \\
27 & Cluster Shade (GLCM) & 0.62 & 0.63 & 0.620 & 1.66 (1.25-2.19) & 9.31e-3 \\
28 & Entropy-PCA2 (MM covariance) & 0.28 & 0.84 & 0.542 & 1.70 (1.22-2.38) & 9.67e-3 \\
29 & Energy (GLCM) & 0.43 & 0.77 & 0.588 & 1.65 (1.22-2.23) & 9.91e-3 \\
30 & SZHGE (GLSZM) & 0.80 & 0.32 & 0.557 & 1.77 (1.32-2.36) & 1.13e-2 \\
31 & Std-PCA1 (MM Granularity) & 0.62 & 0.55 & 0.600 & 1.64 (1.24-2.17) & 0.32e-2 \\
32 & Homogeneity (GLCM) & 0.36 & 0.79 & 0.578 & 1.65 (1.21-2.27) & 1.35e-2 \\
\bottomrule
\end{tabular}
\label{table:proStu}
\end{table}

The tumor volume appears to be an important prognostic feature, as previously published in the radiomics literature~\cite{takenaka2016}~\cite{morgensztern2012}. It appears on the fourth rank in our study with a $p$-value of 4.24e-5 (Se=67\%, Sp=54\%, AUC=0.565 and HR=2.02). In addition to volume, two other first-order classical features are relevant: mean intensity (Se=57\%, Sp=63\%, AUC=0.596 and HR=1.76) and kurtosis (Se=60\%, Sp=65\%, AUC=0.613 and HR=1.73). Finally, 24 classical texture features appear to have a relevant prognostic power (7 from GLCM, 8 from GLRLM, 5 from GLSZM and 4 from GLDM).

Considering the MM features, 5 of them are relevant according to survival analysis: 2 from granularity (Volume-PCA1 and Std-PCA1) and 3 from covariance (Kurtosis-PCA3, Energy-PCA2, Entropy-PCA2). We notice that none of the PCA components higher than the third rank have been selected by the statistical analysis. This means that the components with larger variance contain the relevant information. Volume-PCA1 (MM granularity) is the second most important feature of this study according to the AUC with a value of 0.631, behind strength (GLDM) (AUC=0.633), but ahead of tumor volume (AUC=0.565). Two other relevant MM features have AUC higher than 0.6, Kurtosis-PCA3 (MM Covariance) with an AUC of 0.618 and Std-PCA1 (MM granularity) with an AUC of 0.600.

\begin{figure}[h]
     \centering
     \begin{subfigure}[b]{.4\textwidth}
         \centering
         \includegraphics[width=7cm]{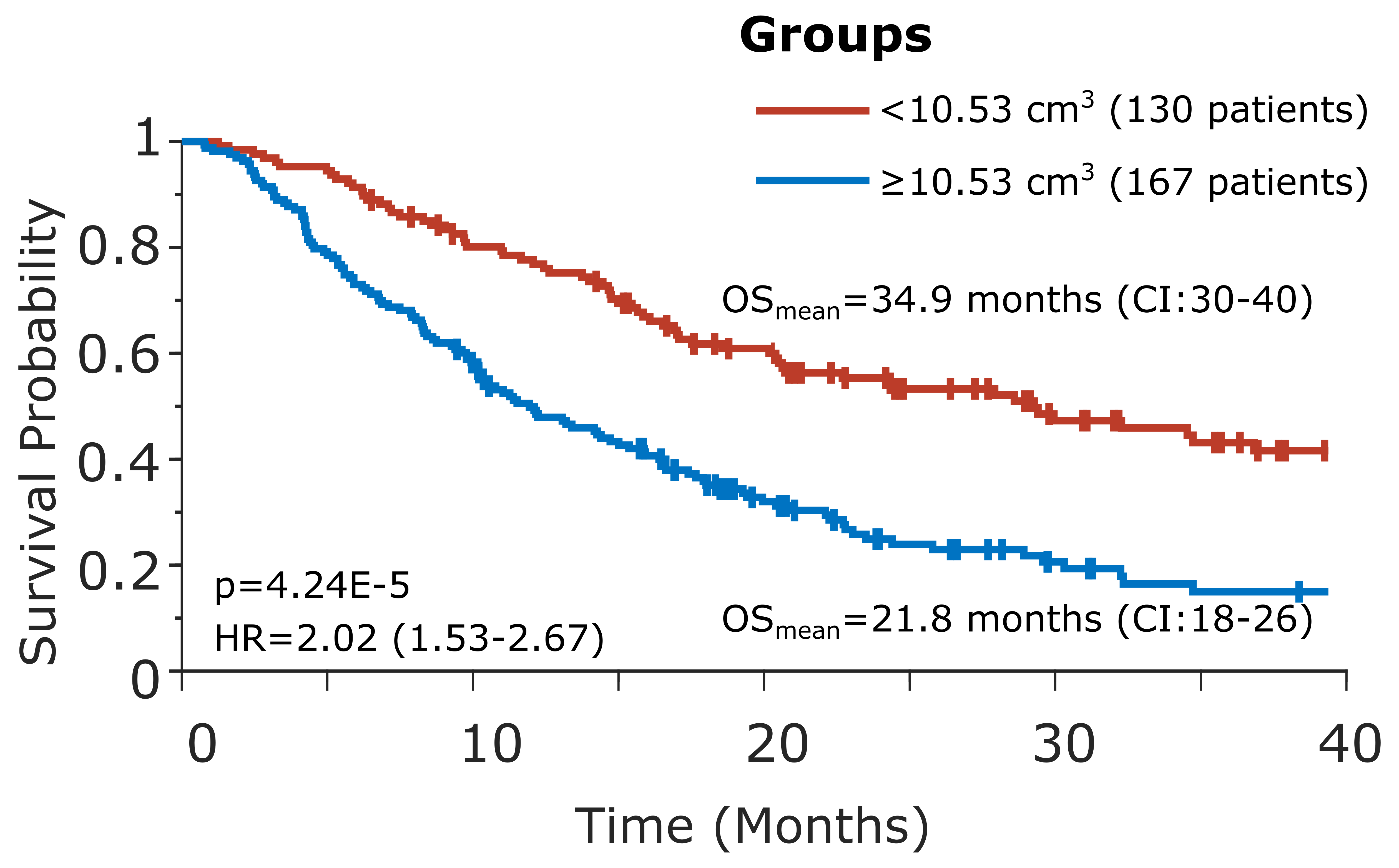}
         \caption{Tumor Volume}
         \label{fig:KMa}
     \end{subfigure}
     \hfill
     \begin{subfigure}[b]{.4\textwidth}
         \centering
         \includegraphics[width=7.5cm]{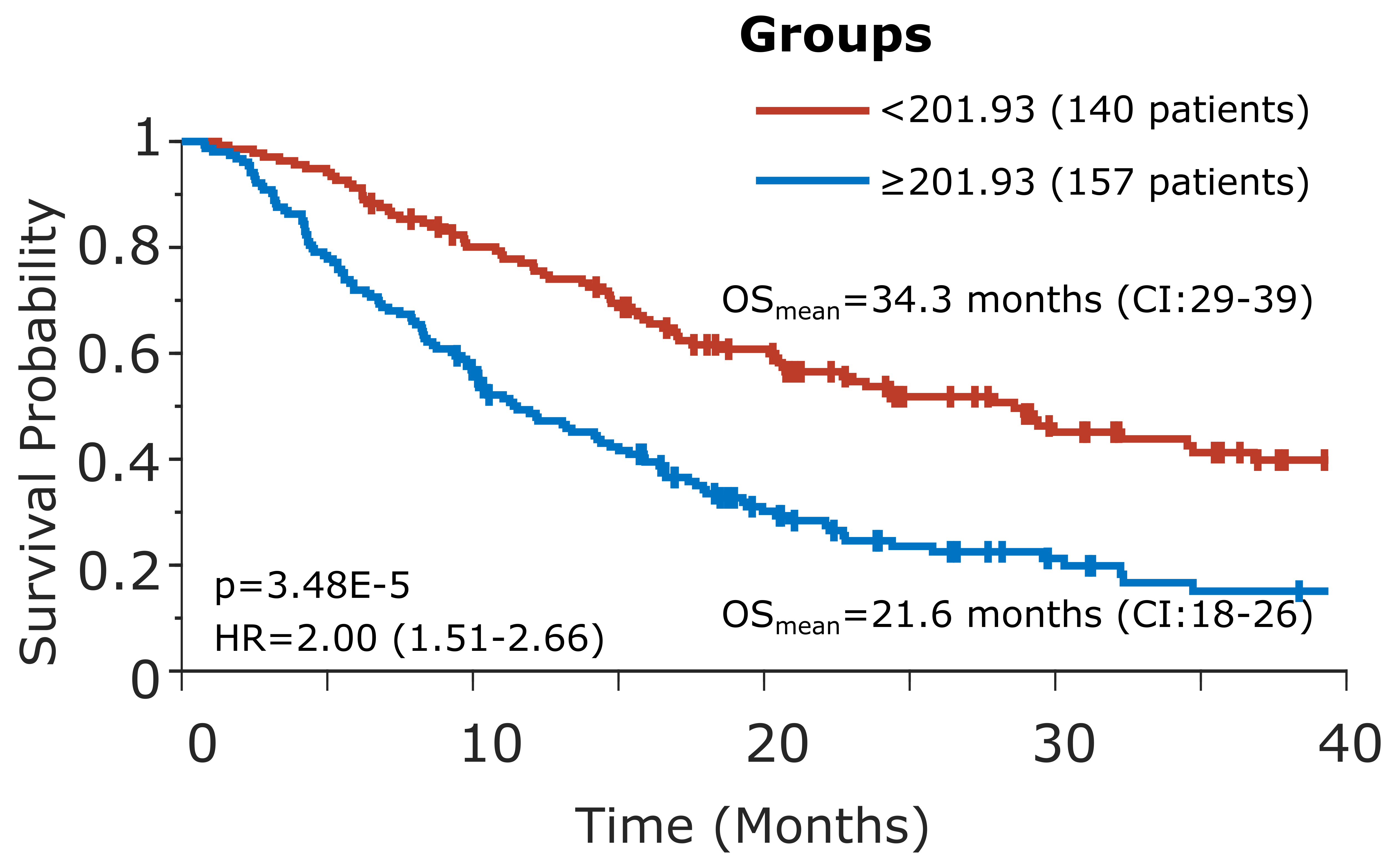}
         \caption{GLNUz (GLSZM)}
         \label{fig:KMb}
     \end{subfigure}
     
     \begin{subfigure}[b]{.4\textwidth}
         \centering
         \includegraphics[width=7cm]{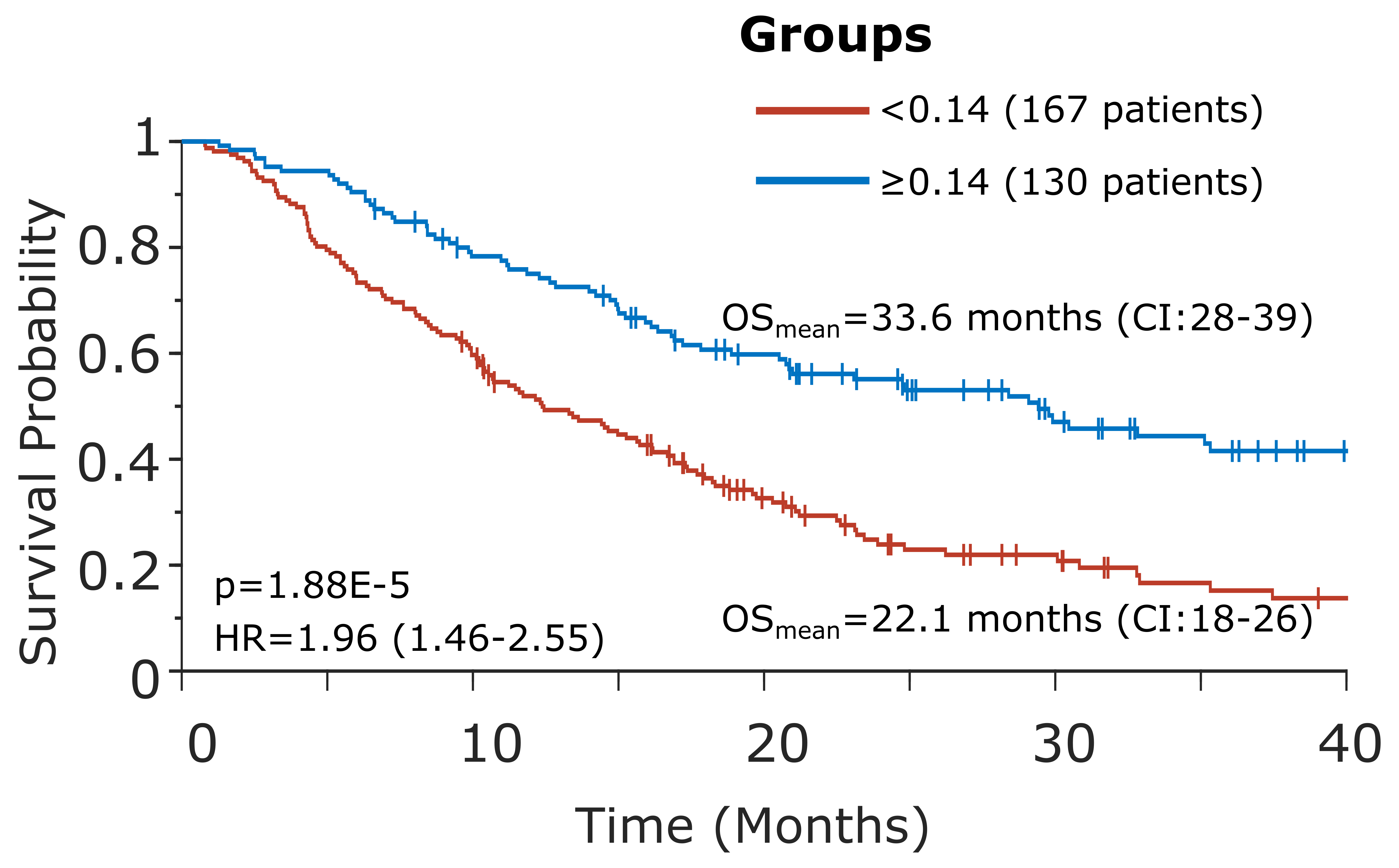}
         \caption{Strength (GLDM)}
         \label{fig:KMc}
     \end{subfigure}
     \hfill
     \begin{subfigure}[b]{.4\textwidth}
         \centering
         \includegraphics[width=7.5cm]{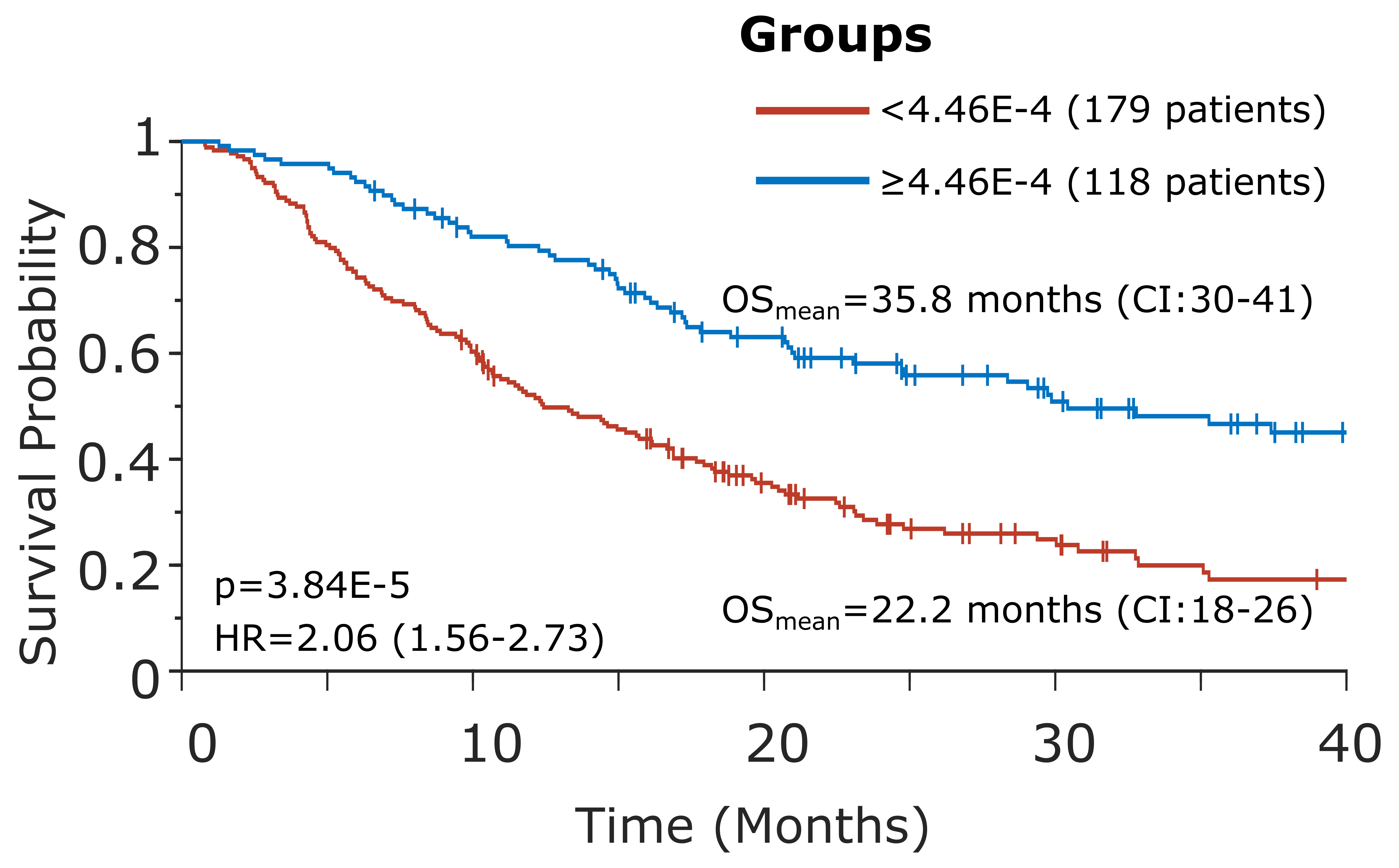}
         \caption{Coarseness (GLDM)}
         \label{fig:KMd}
     \end{subfigure}
     
     \begin{subfigure}[b]{.4\textwidth}
         \centering
         \includegraphics[width=7cm]{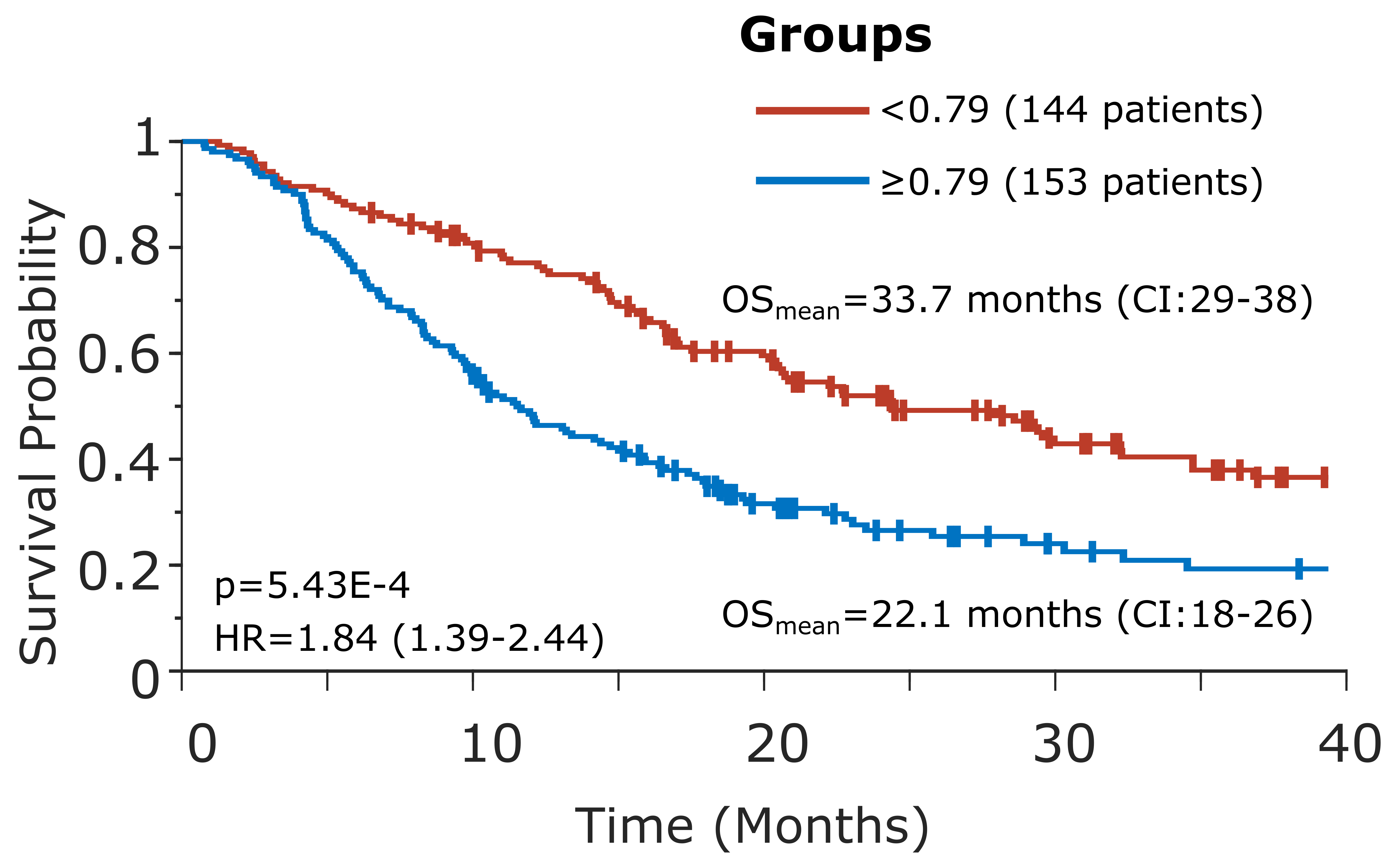}
         \caption{Volume-PCA1 (MM granularity)}
         \label{fig:KMe}
     \end{subfigure}
     \hfill
     \begin{subfigure}[b]{.4\textwidth}
         \centering
         \includegraphics[width=7.5cm]{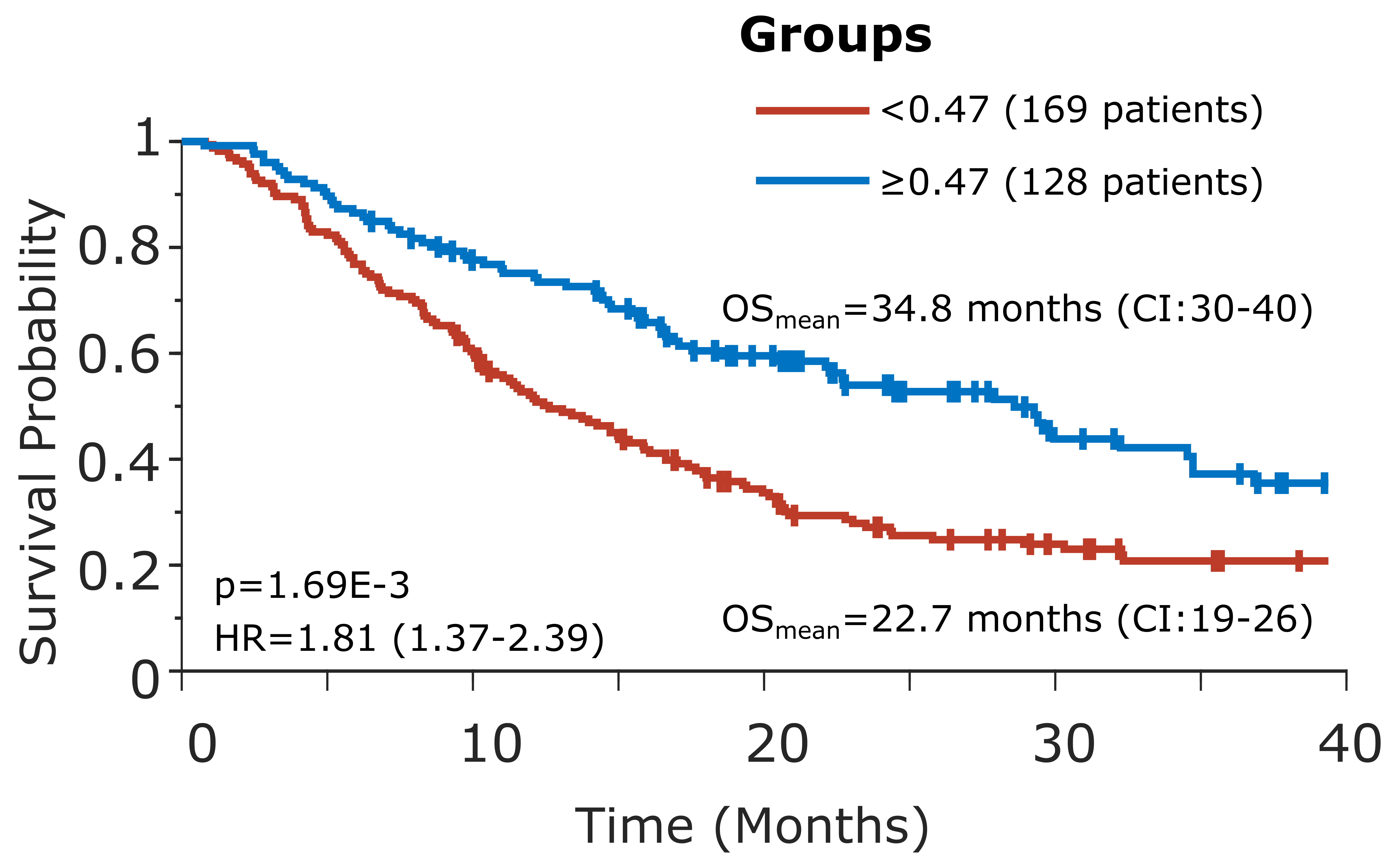}
         \caption{Kurtosis-PCA3 (MM covariance)}
         \label{fig:KMf}
     \end{subfigure}
     
    \caption{Kaplan-Meier survival curves.}
    \label{fig:KM}
\end{figure}

In Figure~\ref{fig:KM} are given the Kaplan-Meier survival curves for six of the most interesting features. it shows that these features are able to separate the cohort into a high-risk and a low-risk group. For instance, patients having a tumor volume higher or equal to 10.53~cm\textsuperscript{3} have 2.02 times more risk to deceased (see Figure~\ref{fig:KMa}). The mean survival time (OS\textsubscript{mean}) for this group is 21.8~months (CI: 18-26) while it is 34.9~months (CI: 30-40) for the patients having a smaller tumor.

This observation is the same for all the relevant features. For instance, volume-PCA1 (MM granularity, Figure~\ref{fig:KMe}) separates two groups of patients: one with an OS\textsubscript{mean}=33.7~months (CI: 29-38) and the other with an OS\textsubscript{mean}=22.1~months (CI: 18-26). The HR shows that you have 1.84 times more risk if you have a high value of this MM feature.

In addition, our experiments did not show any clinical features as relevant: $p$-value equal to 0.0683 for TNM stage and 0.869 for histology (both higher than 5\%). On the other hand, both demographic features (age and gender) are presented as relevant: a cut-off of 68~years leads to a $p$-value of 0.0028 and the gender has a $p$-value equal to 0.0029.

\section{Discussion}
The role of medical imaging in clinical oncology has increased manifold in recent decades. It helps to capture intra-tumoral heterogeneity in a non-invasive way, which the conventional biopsy-based techniques fail to do~\cite{lambin2012}. The framework introduced in this study focused on investigating the prognostic power of MM based features extracted from CT images. By extracting both the CR and MM features, we showed the comparison between the two. To our knowledge, this is the first time MM features are studied in a radiomics analysis.

The CR features (texture based, shape based and first-order features) and MM features (granularity and morphological covariance features) are extracted from an open database of segmented NSCLC patients, captured using CT images~\cite{aerts2015}. To reduce the dimensionality and remove the redundancy of the 1,589 studied features, correlation analysis and PCA are performed. Figure~\ref{fig:correl} shows that several correlations exist between the different iterations of MM features. Hence, a PCA is performed on these features to get transformed features which are much less correlated and bring out useful information. PCA helps us to reduce the number of total features from 1,589 to 159 (49 classical radiomics features and 110 MM features). Although some work has been published on the number of PCA components that should be retained, researchers use a combination of the published rules since the results vary with the dataset used~\cite{jackson1993}~\cite{saccenti2015}~\cite{josse2012}. In this study, we retained 5 PCA components since they contained most of the information.

To analyze the prognostic power of features a Kaplan-Meier survival analysis is performed. We choose to set the horizon at two years of survival because half of the patients were deceased at this stage, making the data balanced. We found that among the 159 features studied, 32 are relevant with a $p$-value inferior to 5\% after post-hoc corrections. Among these features, 10 have an AUC higher or equal to 0.6, out of which 7 are CR features (with Strength (GLDM) having the highest AUC of 0.633) and 3 are MM features (Volume-PCA1 and Std-PCA1 from granularity and Kurtosis-PCA3 from covariance). Among the 32 relevant features, five are MM features, proving the interest of these features to predict the survival of patients suffering from lung cancer.

The Kaplan-Meier survival curves (see Figure~\ref{fig:KM}) show that relevant features successfully classify the population into high and low risk groups. Through this study, we are able to bring out new image biomarkers based on MM, to be used in survival analysis and as predictors for non invasive personalized medicine. It is also interesting to observe that the clinical features (TNM stage and histology) did not play a role in the analysis with a $p$-value higher to 5\%.

The present study has some limitations. Because we are retrospectively studying an open source database we do not have control on all the parameters. Combining imaging data across sites leads to variations in the radiomics features and their robustness due to the use of different hardware tools, acquisition and reconstruction techniques~\cite{mackin2015}. To remove such unwanted sources of variation while preserving biological associations of interest, data harmonisation techniques are proposed in the literature such as Combat harmonization~\cite{fortin2017}~\cite{fortin2018}. Nevertheless, we were not able to use it because of the lack of knowledge about the acquisition devices used for each patient. Even though the results show that the MM features are relevant on this database, they need to be tested on other databases to establish their relevance in general. Diverse data from different databases and multiple imaging modalities can be used for further testing. We expect that using magnetic resonance imaging and PET instead of CT could bring out better results with the MM features, since these techniques are better suited for texture analysis~\cite{lubner2017}~\cite{lu2016}~\cite{hassan2017}.

\section{Conclusion}
In this study we investigated the prognostic power of MM based features in a radiomics study for NSCLC patients. Among the 1,589 studied features, 32 were found relevant to predict patient survival: 27 classical radiomics features and five MM features (including both granularity and morphological covariance features). These features will contribute towards the prognostic models, and eventually to clinical decision making and the course of treatment for patients. 

\bibliographystyle{unsrt}  

\end{document}